# Software Oriented Data Monitoring System

Phani Nandan K. and Pavan Kumar K.

One area of biomedical instrumentation that is becoming increasingly familiar to the general public is that of patient monitoring. Here electronic equipment provides a continuous watch over the vital characteristics and parameters of the critically ill. In the coronary care and the intensive care units in the hospitals, thousands of lives have been saved in the recent years because of the careful and accurate monitoring afforded by this equipment.

Essentially, patients are monitored because they have an unbalance in their body systems. This can be caused by a heart attack or stroke, which can drastically disturb these systems. By continual monitoring, the patient problems can be detected as they occur and remedies taken before these problems get out of hand.

The need for intensive care and patient monitoring has been recognized for centuries. The 24-hour nurses for the critically ill patient has, over the years, become a familiar part of the hospital scene. Nurses are still there but roles have changed somewhat, for they now have powerful tools at their disposal for acquiring and assimilating information about the patients under their care. They are therefore able to render better service to a larger number of patients and are better able to react promptly and properly to an emergency situation. With the capability of providing an immediate alarm in the event of certain abnormalities in the behavior of the patients' critical data, monitoring equipment makes it possible to summon a physician or nurse in time to administer emergency aid often before permanent damage may occur.

The 21st century has been the age of computers and information technology. Computers and processors are used in various automation systems with great accuracy and reliability.

The field of telecommunications has also seen an enormous advancement with new technologies being introduced frequently, particularly the mobile phone technology. Automation, precision and accuracy are a must in the medical field and now-a-days machines are replacing the less accurate humans.

Patients having critical conditions are admitted in the Intensive Care Unit (ICU) for monitoring their condition regularly. Different sensors are attached to their bodies and a dedicated person constantly monitors the data. But by making use of advance mobile technology, messages containing the real time parameters of the patient placed in ICU can be monitored and sent to the concerned doctor even if he is busy attending other patients, thus saving his valuable time. A typical system implementing these features would look like the figure below.

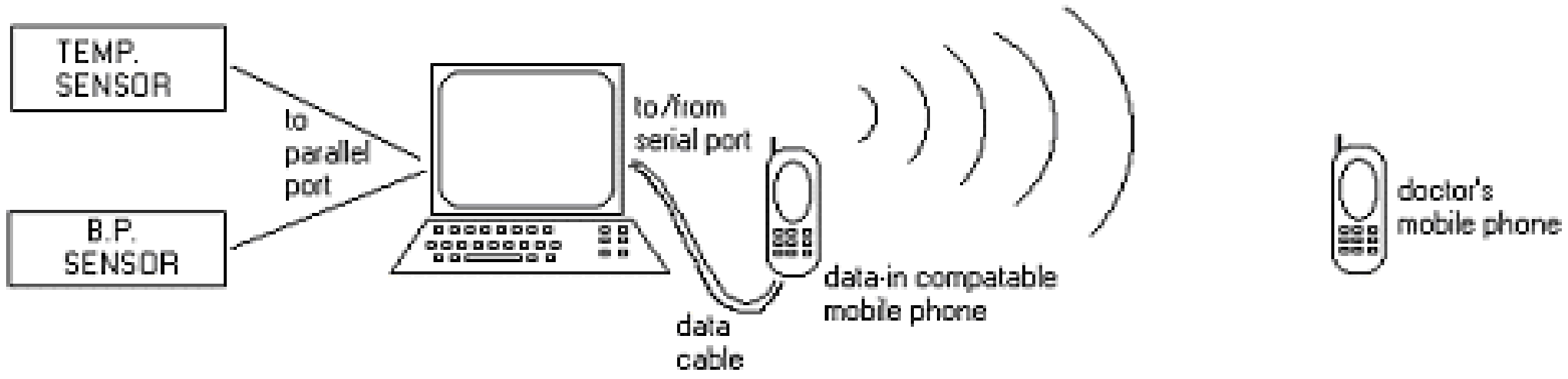

**Fig 1.1 Schematic diagram of *Real time Patients' data Monitoring System***

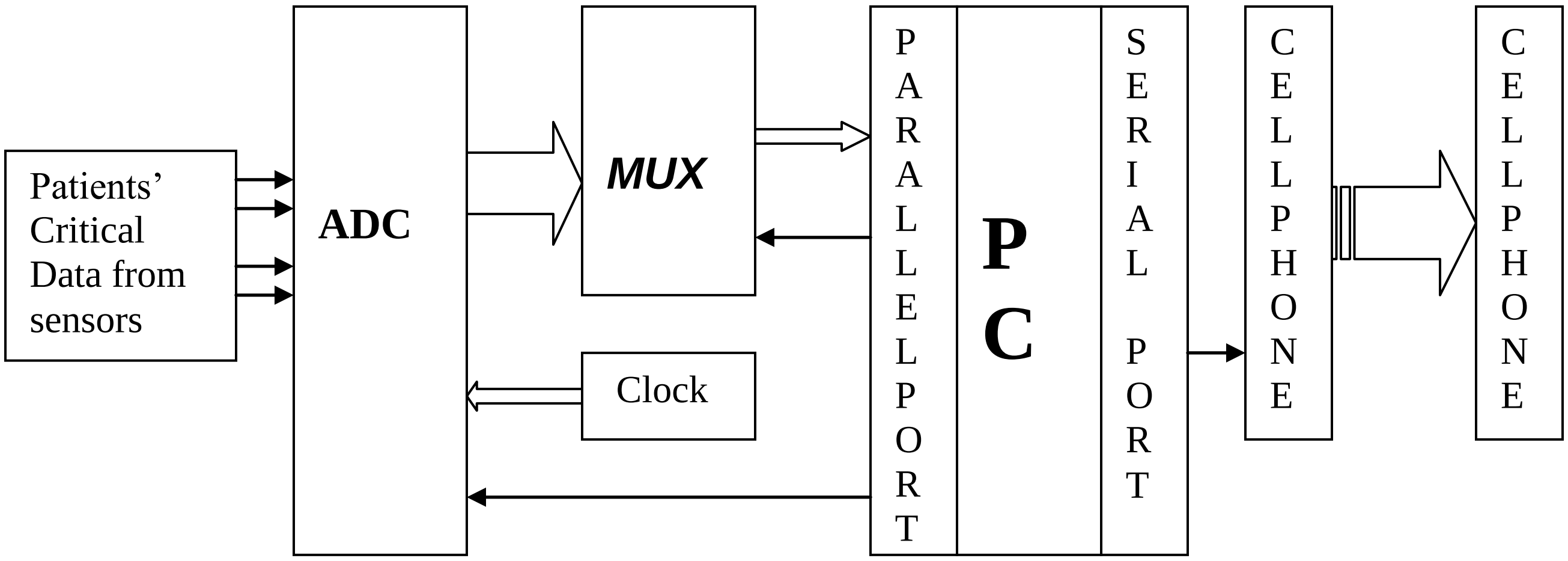

**Fig 1.2 Block diagram of *Real time Patients' data Monitoring System***

The patients' temperature and the heart rate measured by the sensors are sent to the ADC. The ADC consists of 8 input channels. Here the analog inputs i.e. the critical parameters of the patient are converted into digital form and are sent to the computer through the parallel port. Since the parallel port of the computer has only 5 input pins a Multiplexer is used so that a nibble (4 bits) of data is sent at a time. This data in the computer is sent to the Cell phone using serial port of the computer. This cell phone sends SMS to the Doctors cell at appropriate intervals.

## 1.1 CHARACTERISTICS OF MEDICAL DATA:

The three basic types of data that must be acquired, manipulated and archived in the hospital are alphanumeric data, images and physiological signals. Alphanumeric data include the patient's name and address, identification number, results of lab test and physician's notes. Images include X-rays and scans from Computed Tomography, Magnetic Resonance Imaging and Ultrasonic imaging. Examples of physiological signals are the electrocardiogram (ECG), the electroencephalogram (EEG) and BP tracing.

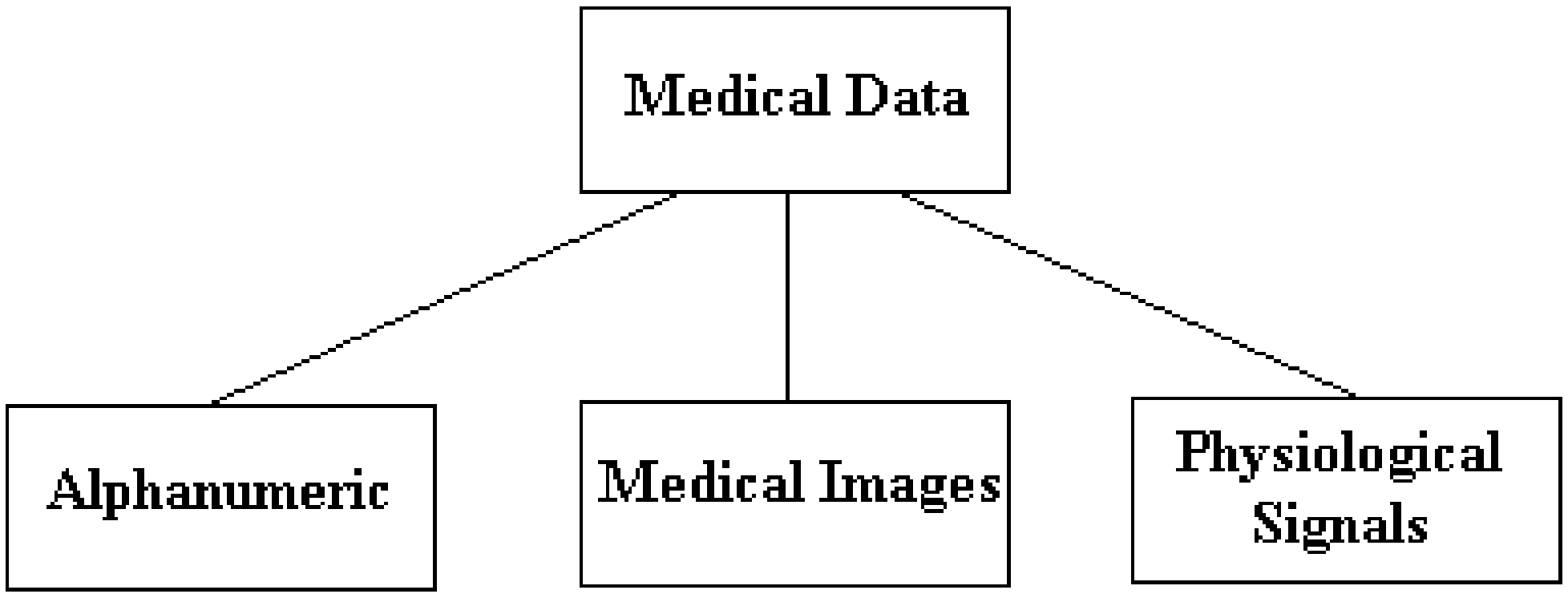

**Fig 1.3 Types of Medical Data**

Quite different systems are necessary to manipulate each of these three types of data. Alphanumeric data are generally managed and organized into a database using a general-purpose mainframe computer.

Image data are traditionally archived on film. However, we are evolving towards picture archiving and communication systems (PACS) that will store images in digitized form on the optical disks and distribute them on demand over a high-speed Local Area Network (LAN) to very high-resolution graphics display monitors located throughout a hospital.

On the other hand, physiological signals that are monitored during surgery in the operating room require real time processing. The clinician must know immediately if the instrument finds abnormal readings as it analyses the continuous data.

## 1.2 BASIC ELEMENTS OF A MEDICAL CARE SYSTEM:

The block diagram below illustrates the operation of the medical care system. Data collection is the starting point in the health care. The clinician asks the patient questions about medical history, records the ECG, and does blood tests and other tests in order to define the patient's problem. Of course medical instruments help in some aspects of this data collection process and even do some processing of data. Ultimately, the clinician analyses the data collected and decides what the basis of the patient's problem is. This decision of diagnosis leads the clinician to prescribe a therapy. Once the therapy is administrated to the patient, the process continues along the closed loop as shown in the figure with more data collection and analysis until patient's problem is rectified.

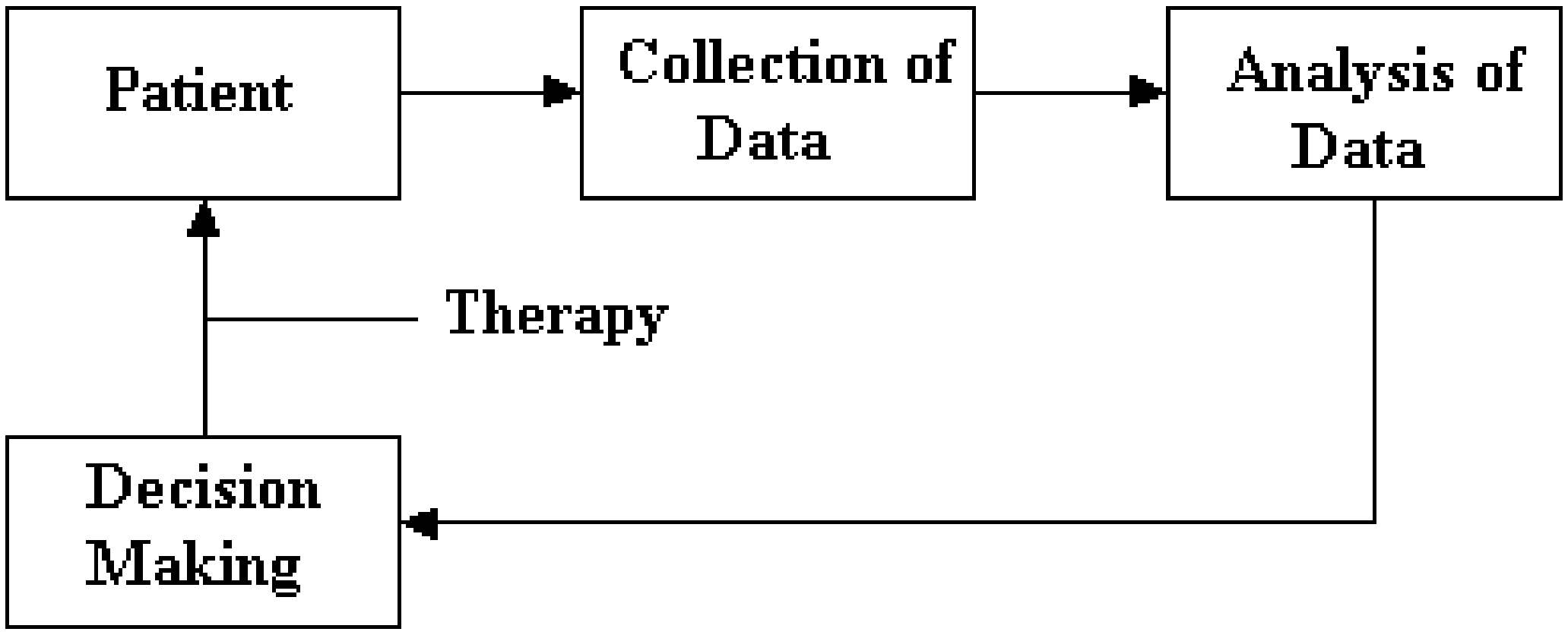

## 2.1 TEMPERATURE SENSOR:

The AD 590 is a two terminal integrated circuit temperature transducer, which produces an output current proportional to absolute temperature.

### 2.1.1 WORKING PRINCIPLE OF AD590:

The AD 590 converts its ambient temperature in degree Kelvin into an output current of 1 µA for every degree Kelvin. In terms of Celsius temperatures, It is 255 µA at 0 F and 310 µA at 100 F. Thus the AD 590 acts as a current source that depends on temperature. The circuit symbol for AD 590 is the same as a current source. The AD 590 is used to build a Celsius thermometer as follows:

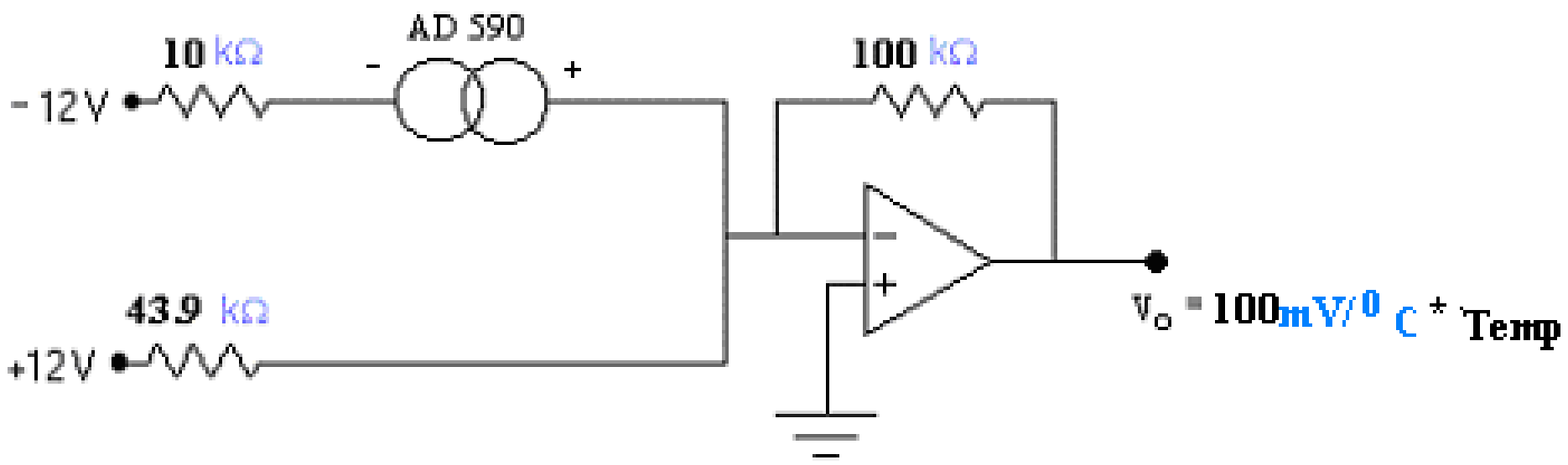

**Fig 2.1: Circuit diagram for Celsius Thermometer using AD 590**

### 2.1.2 CELSIUS THERMOMETER:

In the Celsius thermometer circuit as shown in the figure, all of the AD 590's current is steered into the virtual ground i.e. pin 2 of OpAmp and flows through the 10kΩ feedback resistor, producing a voltage drop equal to $V_o$. Each µA of current thus causes $V_o$ to go positive by (1µA x 100kΩ) = 100mV. A change of $1^oC$ causes $I_t$ to change by 1 µA and consequently produces a change in $V_o$ of 10 mV. The temperature to voltage converter thus has a conversion gain of 100 mV/$^oC$.

At 0 $^oC$, $I_t$ =273µA. But we want $V_o$ to equal 0V. For this reason, an equal and opposite current of 273 µA through 12 V supply and 43.9 kΩ resistor is required. Thus current $I_{Rf}$ flows towards the non-inverting input terminal as it is virtually grounded. To get difference in the current $I_T$ and $I_{Rf}$ , -12V is applied at 10 kΩ which is in series with AD 590, to make the current $I_t$ flow in a direction opposite to that of $I_{Rf}$ . This difference current flows through the feedback resistor, which is proportional to the change in temperature. This results in the net current through $R_f$ to be zero at 0 and thus Vo to be zero volts. For every increase of 1µA/ $^oC$ and above $0^oC$ the net current through $R_f$ increases by 1µA and $V_o$ increases by 100 mV.

## 2.2 HEART RATE SENSOR:

The electrocardiograph or ECG machine permits deduction of many electrical and mechanical defects of the heart by measuring the ECG, which gives the potentials of the heart measured on the body surface. With an ECG machine, one can determine the heart rate and other cardiac parameters. The ECG for one heartbeat has typical values of amplitude and time durations. By measuring them one can determine the heart rate.

The heart rate sensor requires expensive hardware, which is same for ECG and complex conversion algorithms. As the technique used to derive heart rate from an ECG signal is quite difficult, a heart rate sensor has been realized using a simple potentiometer that gives the required analog voltage levels corresponding to different heart rates.

DATA CONVERSION AND PARALLEL PORT INTERFACE

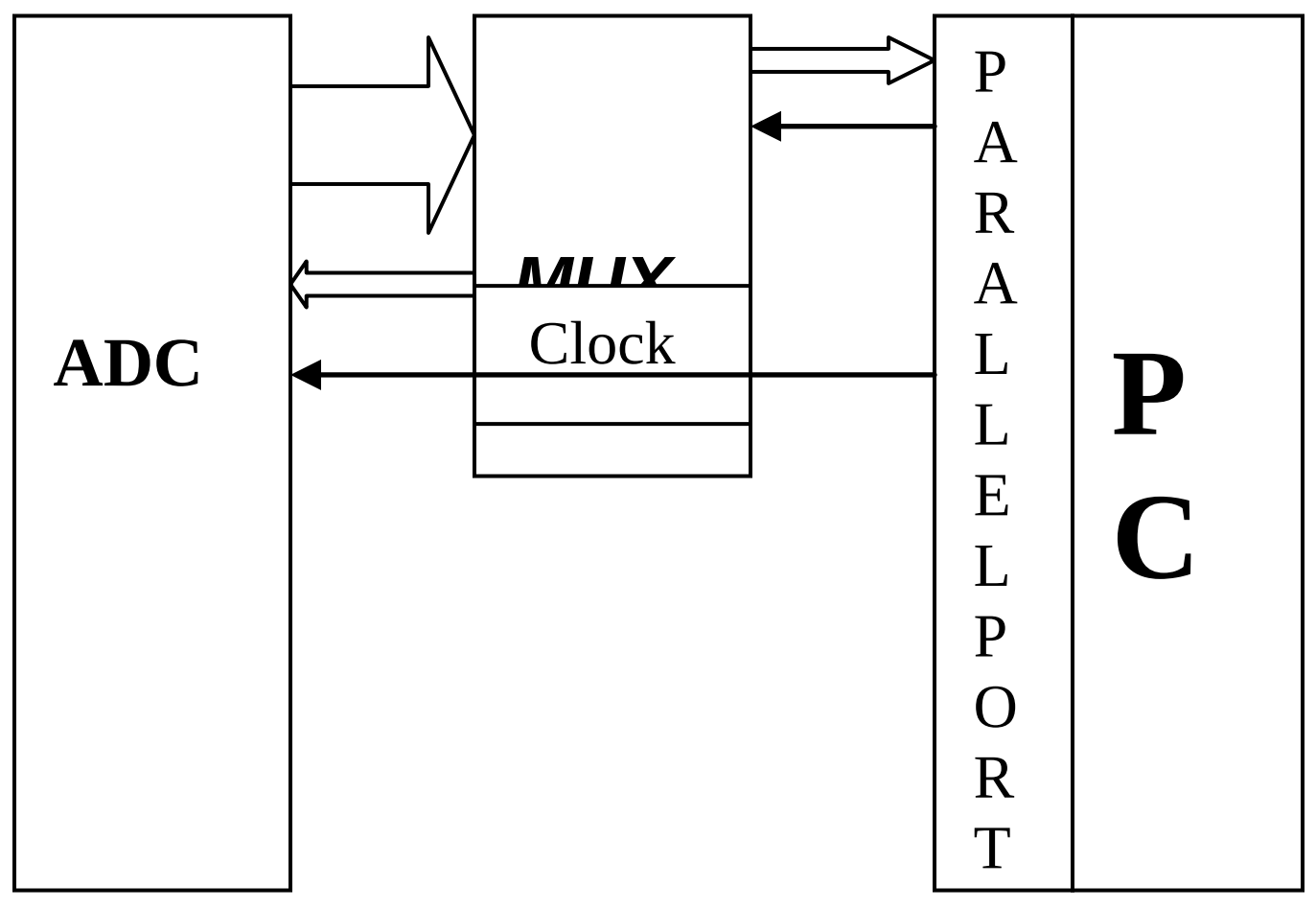

This chapter summarizes the conversion of analog data obtained from the sensor to digital form and also deals with the parallel port interface used to feed data to computer terminal.

## 3.1 ANALOG TO DIGITAL CONVERSION:

The outputs from the sensors are analog but the data to be fed to the computer has to be in the digital form. Hence this is achieved using the Analog to Digital (A/D) converter, ADC 0809 (Refer Fig 3.1 & Fig 3.2).

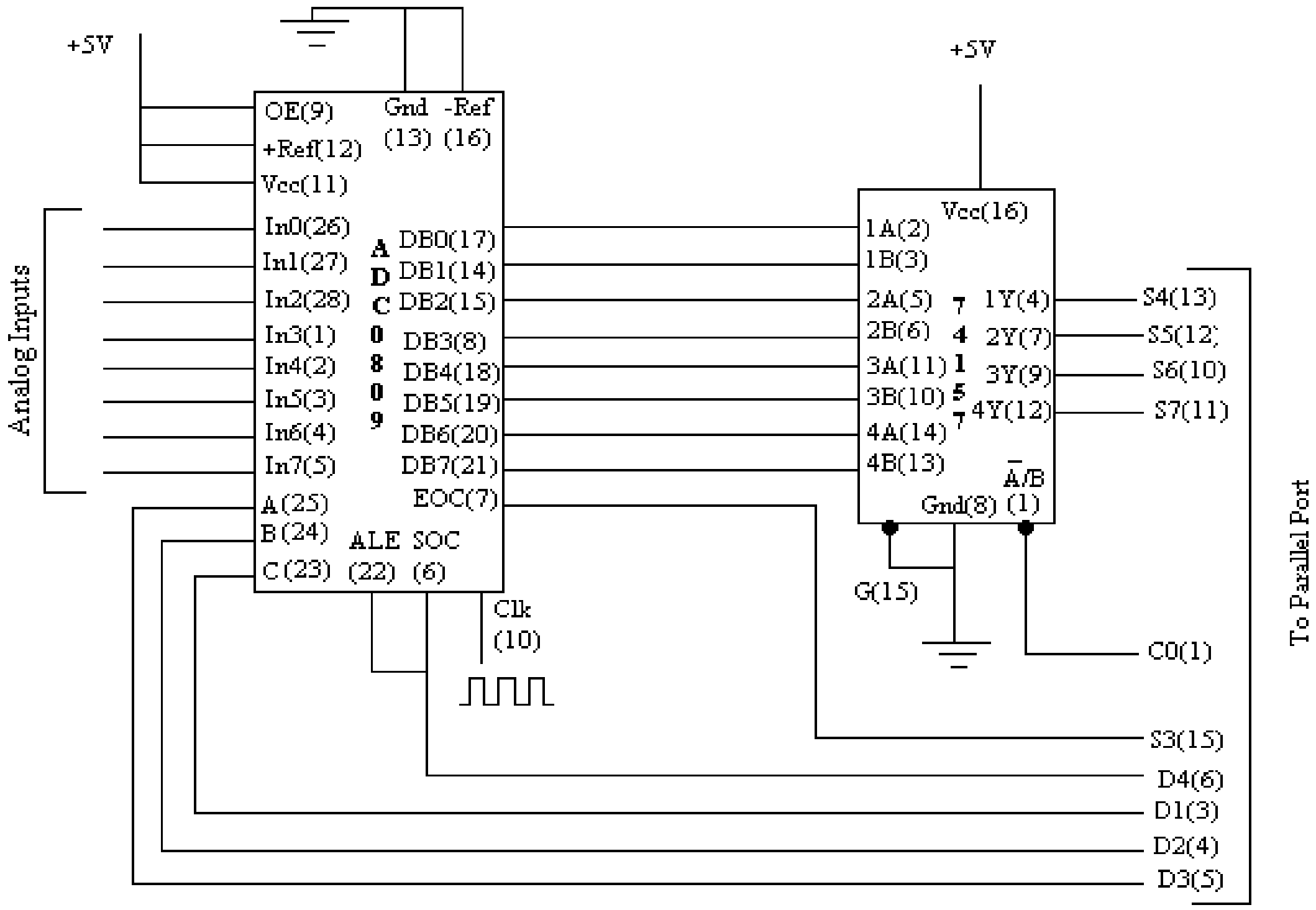

**Fig 3.1: Circuit diagram of ADC along with parallel port interface**

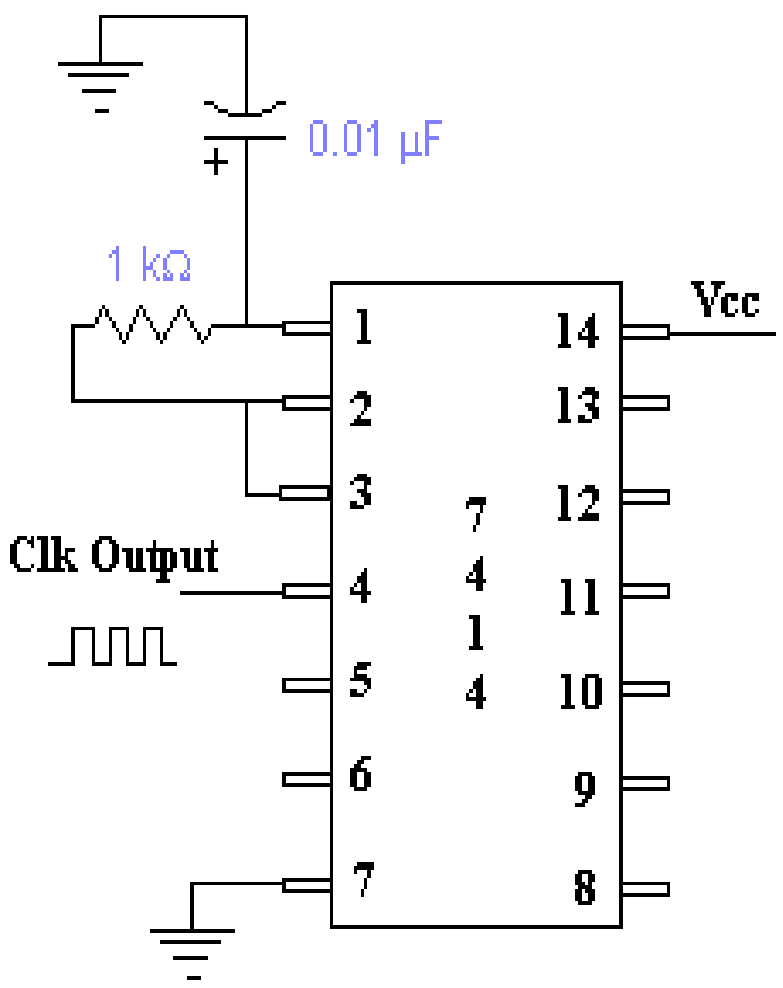

**Fig 3.2 Clock generation using IC 74LS14**

ADC 0808/0809 has 8 input channels and in order to select the desired input channel, it is necessary to send 3-bit address on A, B, C inputs. The address of the desired channel is sent to the Multiplexer address inputs (of the ADC) through port pins. After at least 50 ns this address must be latched. This can be achieved by sending an ALE signal. After another 2.5 µs, the start of conversion (SOC) signal must be sent high and then low to start the conversion process. To indicate end of conversion process ADC 0808/0809 activates EOC signal. The microprocessor system can read converted digital word through data bus by enabling the output enable signal after EOC is activated. The output from ADC is 8-bit binary data. The parallel port (LPT1) of computer has only 5 input pins. For this purpose, a Quad 2X1 Multiplexer (74157) is used. This sends a nibble (4-bits) of data at a time.

## 3.2 THE PARALLEL PORT INTERFACE:

The Parallel Printer Port has 12 digital outputs and 5 digital inputs accessed via 3 consecutive 8-bit ports in the processor's I/O space.

1. 8 output pins accessed via the **DATA Port**
2. 5 input pins (one inverted) accessed via the **STATUS Port**
3. 4 output pins (three inverted) accessed via the **CONTROL Port**
4. The remaining 8 pins are grounded

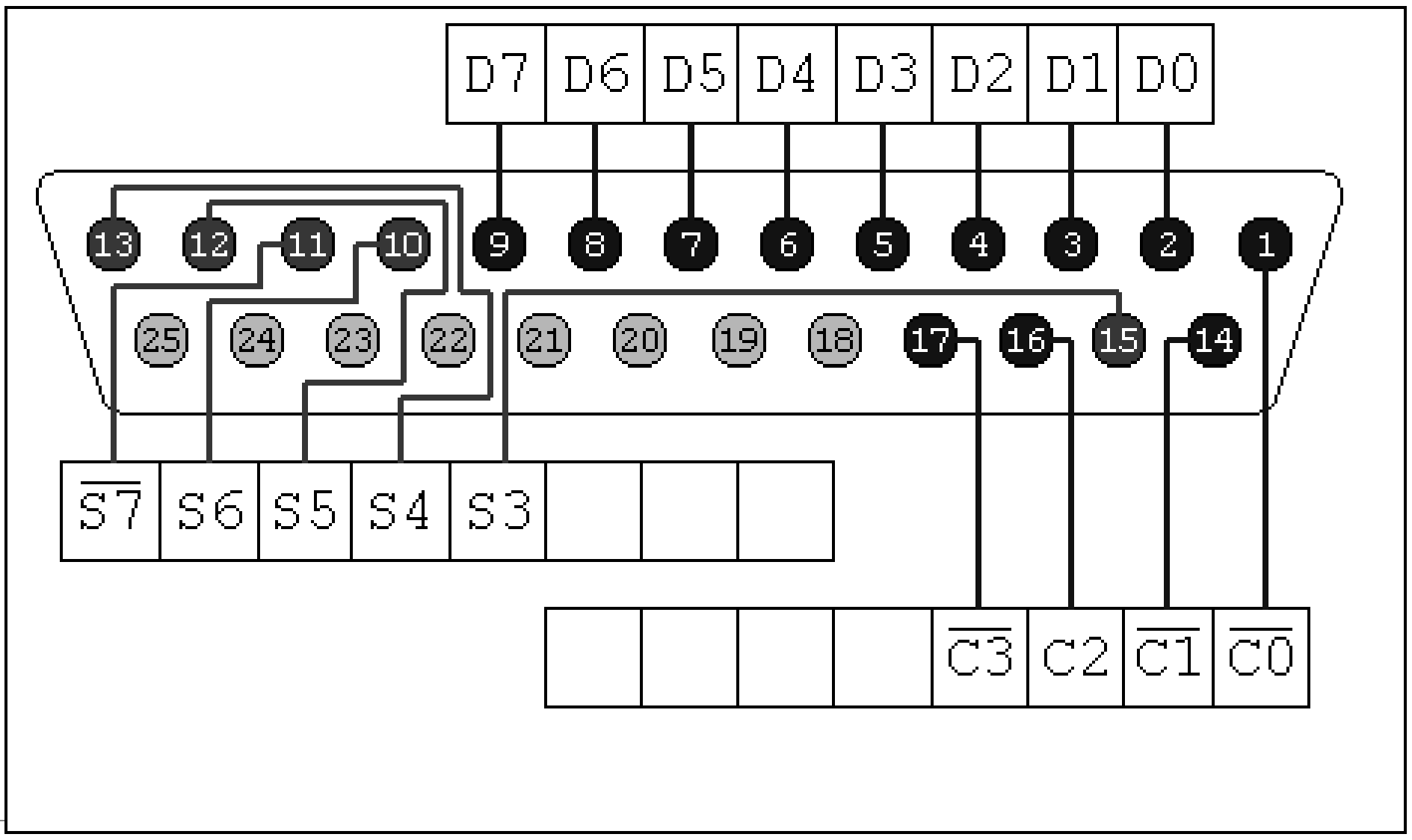

**Fig 3.3: 25-way Female D-Type Connector**

## 3.2.1 PORT ADDRESSES:

The Parallel Port has three commonly used base addresses. These are listed in table 2, below. The 3BCh base address was originally introduced, used for Parallel Ports on early Video Cards. This address then disappeared for a while, when Parallel Ports were later removed from Video Cards. They have now reappeared as an option for Parallel Ports integrated onto motherboards, upon which their configuration can be changed using BIOS.

LPT1 is normally assigned base address 378h, while LPT2 is assigned 278h. However this may not always be the case (as explained later). 378h & 278h have always been commonly used for Parallel Ports. The lower case h denotes that it is in hexadecimal. These addresses may change from machine to machine. Usually, the address is 378h.

| Address | Notes |
|---|---|
| 3BCh - 3BFh | Used for Parallel Ports which were incorporated on to Video Cards - Doesn't support ECP addresses |
| 378h – 37Fh | Usual Address For LPT 1 |
| 278h – 27Fh | Usual Address For LPT 2 |

**Table3.1 Port Addresses**

However to find the address of LPT1 or any of the Line Printer Devices, one can use a lookup table provided by BIOS. When BIOS assigns addresses to your printer devices, it stores the address at specific locations in memory, so we can find them.

| Start Address | Function |
|---|---|
| 0000:0408 | LPT1's Base Address |
| 0000:040A | LPT2's Base Address |
| 0000:040C | LPT3's Base Address |
| 0000:040E | LPT4's Base Address (Note 1) |

**Table3.2 LPT Addresses in the BIOS Data Are**

## 3.2.2 STANDARD PARALLEL PORT (SPP):

| Offset | Name | Read/Write | Bit No | Properties |
|---|---|---|---|---|
| Base + 0 | Data Port | Write (Note-1) | Bit 7 | Data 7 |
| | | | Bit 6 | Data 6 |
| | | | Bit 5 | Data 5 |
| | | | Bit 4 | Data 4 |
| | | | Bit 3 | Data 3 |
| | | | Bit 2 | Data 2 |
| | | | Bit 1 | Data 1 |
| | | | Bit 0 | Data 0 |

**Table3.3 Data Port**

The base address, usually called the Data Port or Data Register is simply used for outputting data on the Parallel Port's data lines (Pins 2-9). This register is normally a write only port. If read from the port, one gets the last byte sent. However if the port is bi-directional, one can receive data on this address (Refer table 3.3).

The Status Port (base address + 1) is a read only port. Any data written to this port will be ignored. The Status Port is made up of 5 input lines (Pins 10,11,12,13 & 15), an IRQ status register and two reserved bits. Bit 7 (Busy) is a active low input (Refer table 3.4).

| Offset | Name | Read/Write | Bit No | Properties |
|---|---|---|---|---|
| Base + 1 | Status Port | Read Only | Bit 7 | Busy |
| | | | Bit 6 | Ack |
| | | | Bit 5 | Paper Out |
| | | | Bit 4 | Select In |
| | | | Bit 3 | Error |
| | | | Bit 2 | IRQ (Not) |
| | | | Bit 1 | Reserved |
| | | | Bit 0 | Reserved |

**Table 3.4 Status Port**

| Offset | Name | Read/Write | Bit No | Properties |
|---|---|---|---|---|
| Base + 2 | Control Port | Read/Write | Bit 7 | Unused |
| | | | Bit 6 | Unused |
| | | | Bit 5 | Enable Bi-Directional Port |
| | | | Bit 4 | Enable IRQ Via Ack Line |
| | | | Bit 3 | Select Printer |
| | | | Bit 2 | Initialize Printer (Reset) |
| | | | Bit 1 | Auto Linefeed |
| | | | Bit 0 | Strobe |

**Table 3.5 Control Port**

The Control Port (base address + 2) is a write only port The control signals used are Strobe, Auto Linefeed, Initialize and Select Printer, all of which are inverted except Initialize (Refer table 3.5).

Normally the Printer Card will have internal pull-up resistors. But some may just have open collector outputs, while others may even have normal totem pole outputs. In order to make device work correctly on as many Printer Ports as possible, one can use an external resistor as well. If there is already an internal resistor, then it will act in Parallel with it, else if it is Totem pole outputs, the resistor will act as a load.

Bits 4 & 5 are internal controls. Bit four will enable the IRQ and Bit 5 will enable the bi-directional port meaning that one can input 8 bits using (DATA 0-7). This mode is only possible if the card supports it. Bits 6 & 7 are reserved. Any writes to these two bits will be ignored.

Some control ports are not open collector, but have totem pole outputs. Therefore, in the interest of portability it is recommend inputting four bits at a time using Multiplexer, reading a nibble at a time.

The connection between parallel port and Multiplexer and ADC are as in the table 3.6. The channels of ADC are selected using address lines A, B, C that are connected to the data port pins D1, D2, D3. The Ale signal is shorted with the SOC and is given to pin D4.

The EOC signal is sensed at the status pin S3. The output bits from the Multiplexer are fed to the status port pins $S7^1$, S6, S5, and S4 of the parallel port. As the $8^{th}$ bit of status port is an inverted input, the bit to be fed is to be complemented. This is done using IC 7414.
The multiplexer select line $A^1$/B is connected to pin $C0^1$ of the control port. When this signal is LOW, higher nibble is selected and when it is HIGH, lower nibble is selected. Thus obtained data is processed accordingly to get the required parameters.

| **Signal** | **Component** | **Parallel Port Pin** | **Port** |
|---|---|---|---|
| A | ADC | 5(D3) | Data Port |
| B | | 4(D2) | |
| C | | 3(D1) | |
| ALE/SOC | | 6(D4) | |
| EOC | | 15(S3) | Status Port |
| $1Y^1$ | MULTIPLEXER | $11(S7^1)$ | |
| 2Y | | 10(S6) | |
| 3Y | | 12(S5) | |
| 4Y | | 13(S4) | |
| $A^1$/B | | $1(C0^1)$ | Control Port |

**Table3.6 Connection between parallel port, multiplexer and ADC**

## 3.2.3 PARALLEL PORT PROGRAMMING:

The aim of the program is to send proper control signals to the ADC and the Multiplexer and read the data from Status Port. It should also convert the data into useful parameters.

As the useful data is in the higher nibble of the status port, to obtain the data the lower nibble is truncated. The lower nibble from the Multiplexer also arrives at the higher nibble of the status port. So it has to be right shifted 4 times to get the data in the lower nibble. Both the nibbles thus obtained are clubbed to get complete 8-bit data. This data is then converted into required parameter using proper conversion techniques. The parameters are checked for safety limits. If they cross the specified limits, then an *Alert SMS* is sent to the concerned doctor. A *Routine SMS* is also sent every 15 minutes even though there is no alert. To send a parameter through SMS, it has to be converted into proper PDU (Protocol Description Unit) format that is discussed in the following sections.

## 4.1 THE GSM TECHNOLOGY:

The acronym GSM stands for Global System for Mobile telecommunications. The digital nature of GSM allows data, both synchronous and asynchronous, to be transported as a bearer service to or from an ISDN terminal. Data can use either the transparent service, which has a fixed delay but no guarantee of data integrity, or a non-transparent service, which guarantees data integrity through an Automatic Repeat Request (ARQ) mechanism, but with a variable delay. The data rates supported by GSM are 300 bps, 600 bps, 1200 bps, 2400 bps, and 9600 bps.

The most basic teleservice supported by GSM is telephony. There is an emergency service, where the nearest emergency service provider is notified by dialing three digits. Group 3 fax, an analog method described in ITUT recommendation T.30, is also supported by use of an appropriate fax adaptor. A unique feature of GSM compared to older analog systems is the *Short Message Service* (SMS). SMS is a bi-directional service for sending short alphanumeric (up to 160 bytes) messages in a store and - forward fashion. For point-to-point SMS, a message can be sent to another subscriber to the service, and an acknowledgement of receipt is provided to the sender. SMS can also be used in a cell broadcast mode, for sending messages such as traffic updates or news updates. Messages can be stored in the SIM card for later retrieval.

Supplementary services are provided on top of teleservices or bearer services, and include features such as caller identification, call forwarding, call waiting, multiparty conversations, and barring of outgoing (international) calls, among others.

## 4.1.1 ARCHITECTURE OF THE GSM NETWORK:

A GSM network is composed of several functional entities, whose functions and interfaces are defined. The GSM network can be divided into three broad parts. **The subscriber carries The Mobile Station**; the **Base Station Subsystem** controls the radio link with the Mobile Station. **The Network Subsystem**, the main part of which is the Mobile services Switching Center, performs the switching of calls between the mobile and other fixed or mobile network users, as well as management of mobile services, such as authentication.

## 4.1.2 MOBILE STATION:

The mobile station (MS) consists of the physical equipment, such as the radio transceiver, display and digital signal processors, and a smart card called the Subscriber Identity Module (SIM).

## 4.1.3 BASE STATION SUBSYSTEM:

The Base Station Subsystem is composed of two parts, the Base Transceiver Station (BTS) and the Base Station Controller (BSC). These communicate across the specified

Abis interface, allowing (as in the rest of the system) operation between components made by different suppliers

## 4.1.4 NETWORK SUBSYSTEM:

The central component of the Network Subsystem is the Mobile services Switching Center (MSC). It acts like a normal switching node of the PSTN or ISDN, and in addition provides all the functionality needed to handle a mobile subscriber, such as registration, authentication, location updating, handovers, and call routing to a roaming subscriber. These services are provided in conjunction with several functional entities, which together form the Network Subsystem. The MSC provides the connection to the public fixed network (PSTN or ISDN), and signaling between functional entities uses the ITUT Signaling System Number 7 (SS7), used in ISDN and widely used in current public networks.

The mobile phones that can be connected to the computer are **Siemens c35i, Nokia-5000 series, Nokia-8000 series** etc. The mobile phone used here is Siemens **c35i**. The mobile is connected to the serial port (COM1) of computer through a data cable.

The GSM mobile telephone can be controlled using AT commands, where AT+C commands according to ETSI GSM 07.07 and GSM 07.05 specification as well as several manufacturer specific AT commands are available.

## 4.2 SMS AND THE PDU FORMAT:

Short Message Service (SMS) is the ability to send and receive short alphanumeric messages to and from mobile telephones. SMS can also be used as a transport for binary payloads and to implement the WAP stack on top of the SMSC bear. SMS was created as part of the GSM Phase 1 standard. SMS allows users to directly transmit messages to each other without the use of an operator (it is, however, necessary to have the underlying operator controlled wireless service). The first user can send a message to a mobile unit, via a direct connect computer. The SMS protocol of messaging is also

"smarter" then standard paging. SMS is a store and forward method therefore, if the end user is not available, the mobile unit is powered off, or the unit is outside a service area, when the unit comes back on line the message will appear. A SMS message can also be sent "certified," where it will notify the message originator of the end user's receipt of themessage.

## 4.2.1 THE PDU FORMAT:

There are two ways of sending and receiving SMS messages:

1. Text mode
2. PDU (protocol description unit) mode.

The text mode (unavailable on some phones) is just an encoding of the bit stream represented by the PDU mode. Alphabets may differ and there are several encoding alternatives when displaying an SMS message. If one reads the message on phone, the phone will choose a proper encoding. An application capable of reading incoming SMS messages can thus use text mode or PDU mode. If text mode is used, the application is bound to (or limited by) the set of preset encoding options. In some cases, that's just not good enough. If PDU mode is used, any encoding can be implemented. Using Text mode or PDU mode depends on the mobile on which one works. For e.g. Siemens c35i works on PDU mode where as Nokia 5000 series works on Text mode. The mode of operation cannot be changed i.e., text mode operations cannot be performed on a Siemens set and vice versa.

## 4.2.2 RECEIVING A MESSAGE IN THE PDU MODE:

The PDU string contains not only the message, but also some information bits. It is all in the form of hexa-decimal *octets* or decimal *semi-octets*. The following string is received by sending the message containing "hellohello" to a destination mobile “9844120647”.

**000100 0A 81 8944216074 0000 0A E8 32 9B FD 46 97 D9 EC 37**

The first 6 octets represent international standard header. The next 2 octets represent the length of the destination mobile number. For e.g., “9844120647” is of 10 digits (0Ah). The following 10 octets represent the destination mobile number in coded form. The coding is done in such a way that 2 successive digits are taken at a time and swapped. So “9844120647” becomes “8944216074”. The next 4 octets (0000h) indicate the start of message bits followed by length of the message string. In the above example “hellohello” has 10 characters (0Ah). This is followed by the PDU format of the message string.

### 4.2.3 CODING 7-BIT DATA (SEPTET'S) INTO OCTET'S:

The message "hellohello" consists of 10 characters, called septets when represented by 7 bits each (ASCII). These septets need to be transformed into octets for the SMS transfer.

The first septet (h) is turned into an octet by adding the rightmost bit of the second septet. This bit is inserted to the left, which yields 1 + 1101000 = 11101000 ("E8"). The rightmost bit of the second character is then consumed, so the second character (septet) needs two bits of the third character to make an 8bit octet. This process goes on and on yielding the following octets (Refer table 4.1).

| Character | ASCII | PDU Conversion Procedure | | | PDU |
|---|---|---|---|---|---|
| **h** | 104 | 1101000 | 1101000 | **1**1101000 | E8 |
| **e** | 101 | 1100101 | 110010**1** | **00**110010 | 32 |
| **l** | 108 | 1101100 | 11011**00** | **100**11011 | 9B |
| **l** | 108 | 1101100 | 1101**100** | **1111**1101 | FD |
| **o** | 111 | 1101111 | 110**1111** | **01000**110 | 46 |
| **h** | 104 | 1101000 | 11**01000** | **100101**11 | 97 |
| **e** | 101 | 1100101 | 1**100101** | **1101100**1 | D9 |
| **l** | 108 | 1101100 | **1101100** | | |
| **l** | 108 | 1101100 | 1101100 | **1**1101100 | EC |
| **o** | 111 | 1101111 | 110111**1** | 1101111 | 37 |

**Table 4.1 Coding 7-bit data into Octets**

The 9 octets from "hellohello" are E8 32 9B FD 46 97 D9 EC 37

In similar manner the required strings can be converted to PDU formats. The parameters from the patients are combined with patient's name and a string is formed. This string, along with doctor’s mobile phone number is sent as parameter to a function which does the PDU format conversion. The converted string is sent through the COM1 port to mobile phone. To send the PDU format string through the serial port, AT commands are used. They are as follows:

AT+CMGS=21

> 0001000A8189442160740000AE8329BFD4697D9EC37 <ctrl-Z>+CMGS: 182

OK

AT specifies that the following is an AT command to the processor.

+CMGS is command to send message. This instruction has to be followed with the length of PDU such that length= (full length – 2)/2. In this example full length is 44. Therefore length is 21. This instruction has to be sent through the serial port. The mobile phone then responds with a ‘>’ sign. Once the response is received PDU string has to be inserted. The PDU string has to be followed with a <ctrl-z> character to send it to mobile phone. If the request is acknowledged and if the message is sent, the mobile phone responds by sending the address of the memory location (+CMGS: 182) of SIM where the outgoing message is stored and an “ok” to the computer else an “error” is sent. The specifications of serial port are discussed in the next section.

INTERFACING THE SERIAL/RS232 PORT

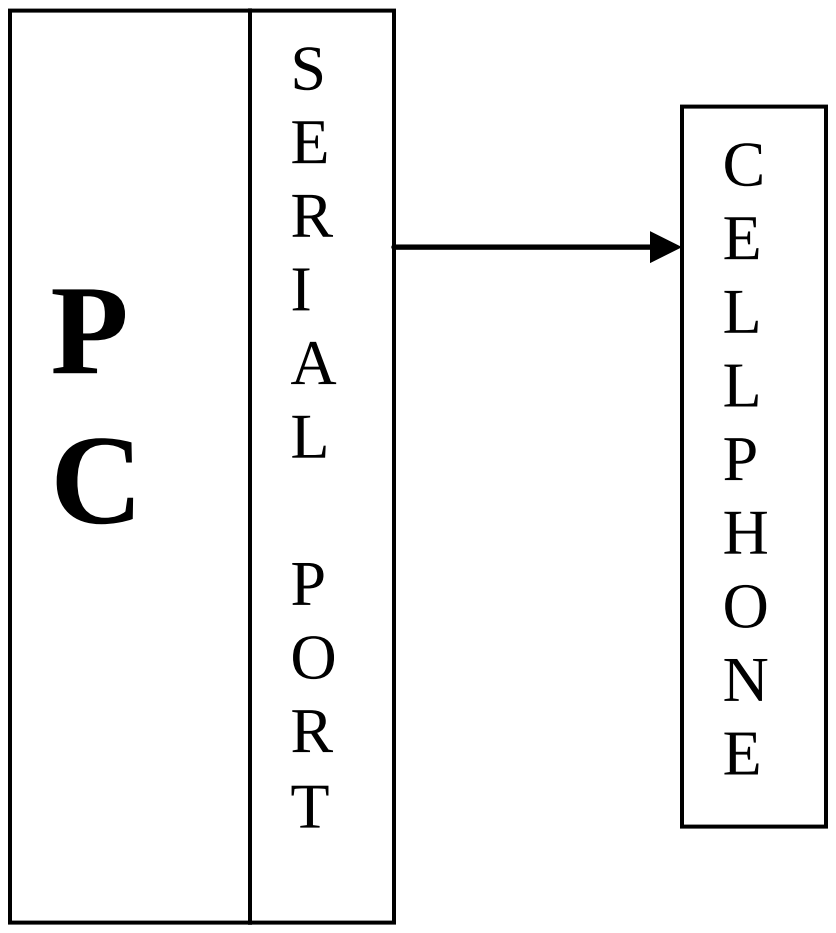

This chapter gives an idea of communicating the mobile phone with the serial port. It deals with the standards and electrical properties of the serial port.

## 5.1 THE SERIAL PORT:

It is necessary to initialize the settings of serial port before using it for any purpose. The settings required to communicate **Siemens c35i** with the serial port are, 19200 bps baud rate, 8 data bits, No parity and 1 stop bit.

### 5.1.1 HARDWARE PROPERTIES:

Devices that use serial cables for their communication are split into two categories. These are DCE (Data Communications Equipment) and DTE (Data Terminal Equipment.) Data Communications Equipment are devices such as your modem, TA adapter, plotter etc while Data Terminal Equipment is your Computer or Terminal.

The electrical specification of the serial port is contained in the EIA (Electronics Industry Association) RS232C standard. It states many parameters such as -

1. A "Space" (logic 0) will be between +3 and +25 Volts.
2. A "Mark" (Logic 1) will be between -3 and -25 Volts.
3. The region between +3 and -3 volts is undefined.
4. An open circuit voltage should never exceed 25 volts. (In Reference to GND)
5. A short circuit current should not exceed 500mA. The driver should be able to handle this without damage. (Take note of this one!)

**Table 5.1 Electrical Properties of RS232C**

Line Capacitance, Maximum Baud Rates etc are also included. It is interesting to note however, that the RS232C standard specifies a maximum baud rate of 20,000 bps !, which is rather slow by today's standards.

Serial Ports come in two "sizes", There are the D-Type 25 pin connector and the D-Type 9 pin connector both of which are male on the back of the PC, thus you will require a female connector on your device. Below is a table of pin connections for the 9 pin and 25 pin D-Type connectors.

## 5.1.2 SERIAL PINOUTS (D25 AND D9 CONNECTORS):

| D-Type-25 Pin No. | D-Type-9 Pin No. | Abbreviation | Full Name |
|---|---|---|---|
| Pin 2 | Pin 3 | TD | Transmit Data |
| Pin 3 | Pin 2 | RD | Receive Data |
| Pin 4 | Pin 7 | RTS | Request To Send |
| Pin 5 | Pin 8 | CTS | Clear To Send |
| Pin 6 | Pin 6 | DSR | Data Set Ready |
| Pin 7 | Pin 5 | SG | Signal Ground |
| Pin 8 | Pin 1 | CD | Carrier Detect |
| Pin 20 | Pin 4 | DTR | Data Terminal Ready |
| Pin 22 | Pin 9 | RI | Ring Indicator |

**Table 5.2 D-Type 9 Pin and D Type 25 Pin Connectors**

## 5.1.3 PIN FUNCTIONS:

| Abbreviation | Full Name | Function |
|---|---|---|
| TD | Transmit Data | Serial Data Output (TXD) |
| RD | Receive Data | Serial Data Input (RXD) |
| CTS | Clear to Send | This line indicates that the Modem is ready to exchange data. |
| DCD | Data Carrier Detect | When the modem detects a "Carrier" from the modem at the other end of the phone line, this Line becomes active. |
| DSR | Data Set Ready | This tells the UART that the modem is ready to establish a link. |
| DTR | Data Terminal Ready | This is the opposite to DSR. This tells the Modem that the UART is ready to link. |
| RTS | Request To Send | This line informs the Modem that the UART is ready to exchange data. |
| RI | Ring Indicator | Goes active when modem detects a ringing signal from the PSTN. |

**Table 5.3 Pin function of RS232C**

## 5.2 RS-232 WAVEFORMS:

RS-232 communication is asynchronous. That is a clock signal is not sent with the data. Each word is synchronized using it's start bit, and an internal clock on each side, keeps tabs on the timing.

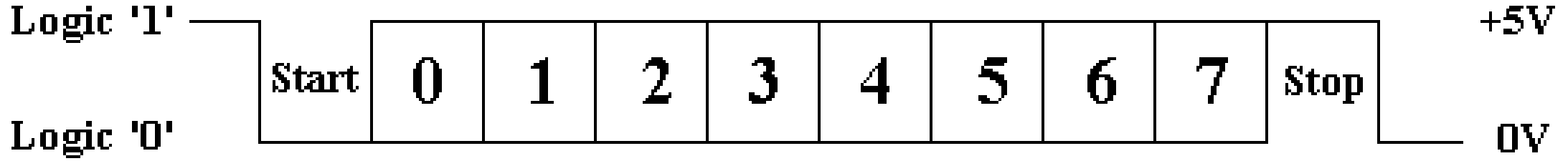

**Fig 5.1 TTL/CMOS Serial Logic Waveform**

The diagram above shows the expected waveform from the UART when using the common 8N1 format. 8N1 signifies 8 Data bits, No Parity and 1 Stop Bit. The RS-232 line, when idle is in the Mark State (Logic 1). A transmission starts with a start bit which is (Logic 0). Then each bit is sent down the line, one at a time. The LSB (Least Significant Bit) is sent first. A Stop Bit (Logic 1) is then appended to the signal to make up the transmission.

The diagram shows the next bit after the Stop Bit to be Logic 0. This must mean another word is following, and this is it's Start Bit. If there is no more data coming then the receive line will stay in it's idle state (logic 1). We have encountered something called a "Break" Signal. This is when the data line is held in a Logic 0 state for a time long enough to send an entire word. Therefore if you don't put the line back into an idle state, then the receiving end will interpret this as a break signal.

The data sent using this method, is said to be *framed*. That is the data is *framed* between a Start and Stop Bit. Should the Stop Bit be received as a Logic 0, then a framing error will occur. This is common, when both sides are communicating at different speeds.

The above diagram is only relevant for the signal immediately at the UART. RS-232 logic levels uses +3 to +25 volts to signify a "Space" (Logic 0) and -3 to -25 volts for a "Mark" (logic 1). Any voltage in between these regions (ie between +3 and -3 Volts) is undefined. Therefore this signal is put through a "RS-232 Level Converter". This is the signal present on the RS-232 Port of your computer, shown below.

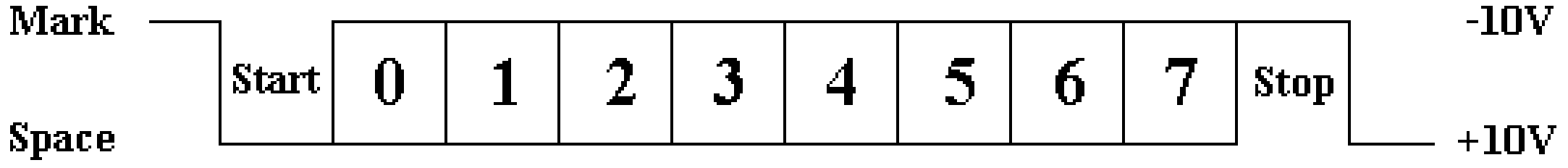

**Fig 5.2 RS-232 Logic Waveform**

The above waveform applies to the Transmit and Receive lines on the RS-232 port. These lines carry serial data, hence the name Serial Port. There are other lines on the RS-232 port, which, in essence are *Parallel* lines. These lines (RTS, CTS, DCD, DSR, DTR, RTS and RI) are also at RS-232 Logic Levels.

The standard port addresses are as follows:

| Name | Address |
|---|---|
| COM 1 | 3F8 |
| COM 2 | 2F8 |
| COM 3 | 3E8 |
| COM 4 | 2E8 |

**Table 5.4 Standard Port Addresses**

# SOFTWARE IMPLEMENTATION

This chapter deals with the software implementation using VB. It also gives a brief idea about the API calls and algorithm.

The programming language used is Visual Basic (VB). VB is a front-end language. So it is not flexible to control the peripherals like communication ports through VB programming. MSCOMM control is used to send or receive data through serial port. To communicate with the parallel port, a win95io.dll file (API) has to be included in System files of Windows.

Windows API calls are what applications use to request services (such as screen control, printers, and memory) from the operating system. An API call in C, Visual Basic, or other languages places a series of values (parameters) at a location in memory (the stack) and then requests the operating system or DLL to execute a function (the procedure call) using the values provided. The function reads the values (call stack) and executes its function code using those values or the data that the values point to. If a result is returned, it is placed at another location (return register) for the calling application to use. This is shown Figure 6.1.

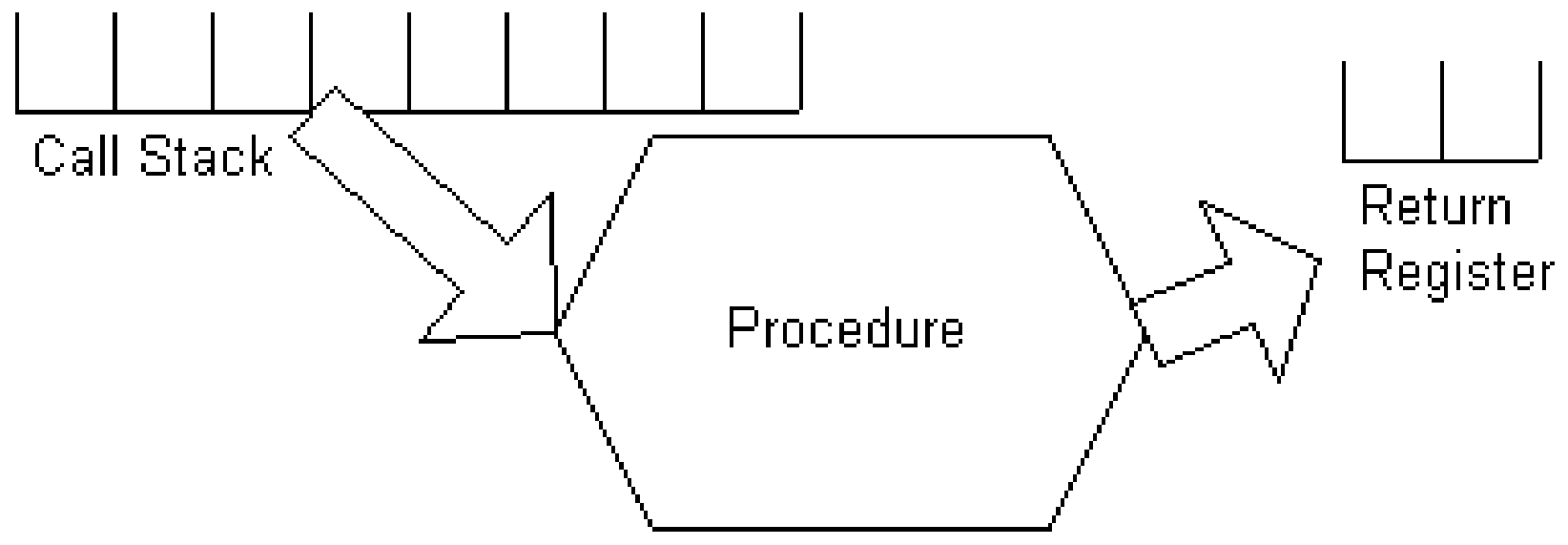

**Fig 6.1 A model of an API call**

## 6.1 ALGORITHM:

1. Initialize A, B, C to 0
2. Select the multiplexer channel by sending proper address to the ADC
3. Latch this address and start the A/D conversion (make ALE/SOC HIGH).
4. Stop the conversion (make ALE/SOC LOW)
5. Check for EOC.
6. If LOW goto step 4 else continue.
7. Read lower nibble followed by higher nibble by sending proper mux select signals
8. Convert the data read to appropriate form.
9. Compare this data with the safe limits.
10. If they have crossed the limit, goto step 13 (send an alert SMS to the concerned doctor), else continue.
11. Check for completion of 15 min and if time out goto step 13 (send routine SMS) else continue
12. Increment A, B, C till all channel are monitored and goto step 1.
13. Create a message string consisting of type of the message (alert or routine), patient's name and doctor's mobile phone number.
14. If the message is a routine signal, check for completion of 5 min since the last SMS was sent and if not continue.
15. Send SMS and return

## 6.2 FLOW CHART:

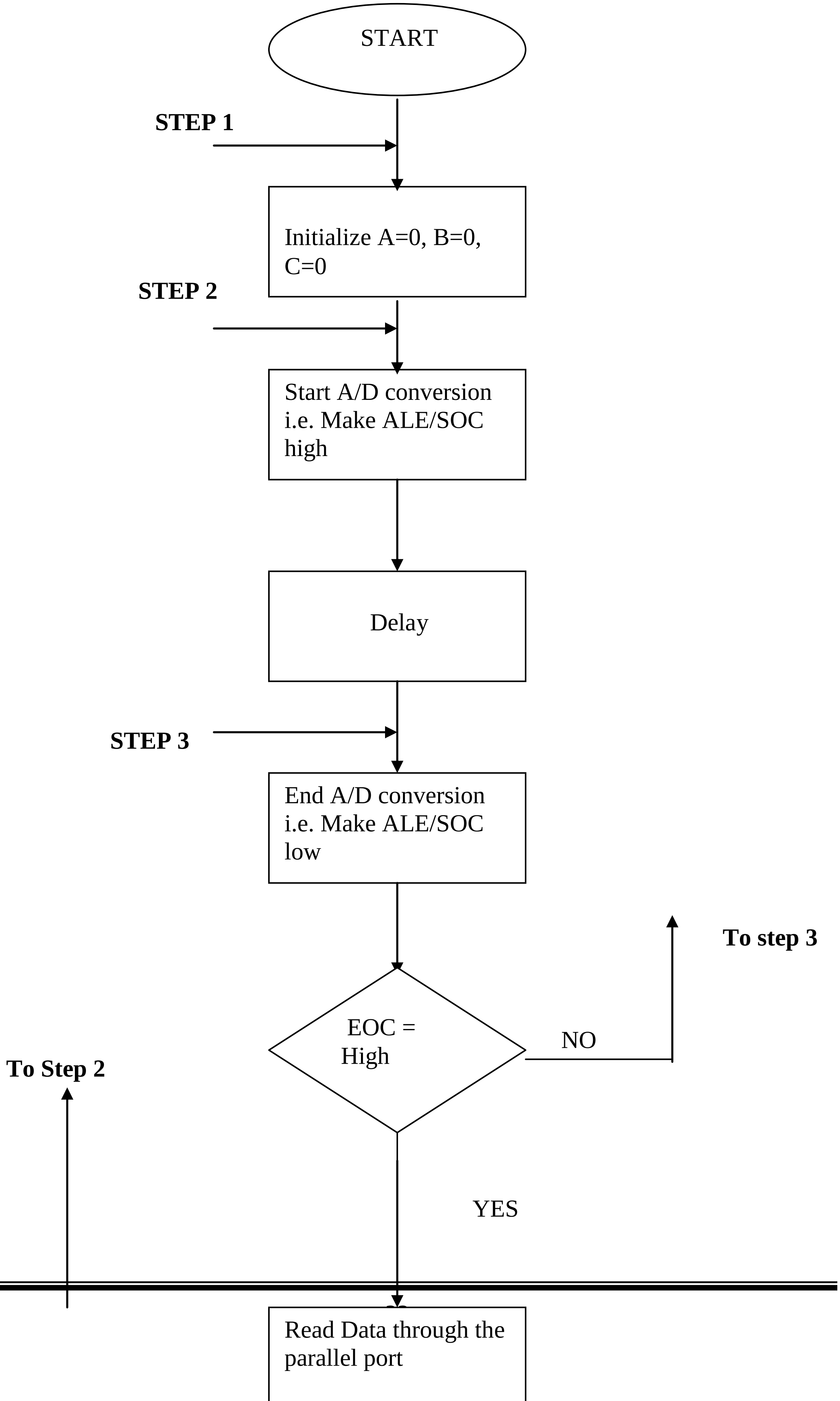

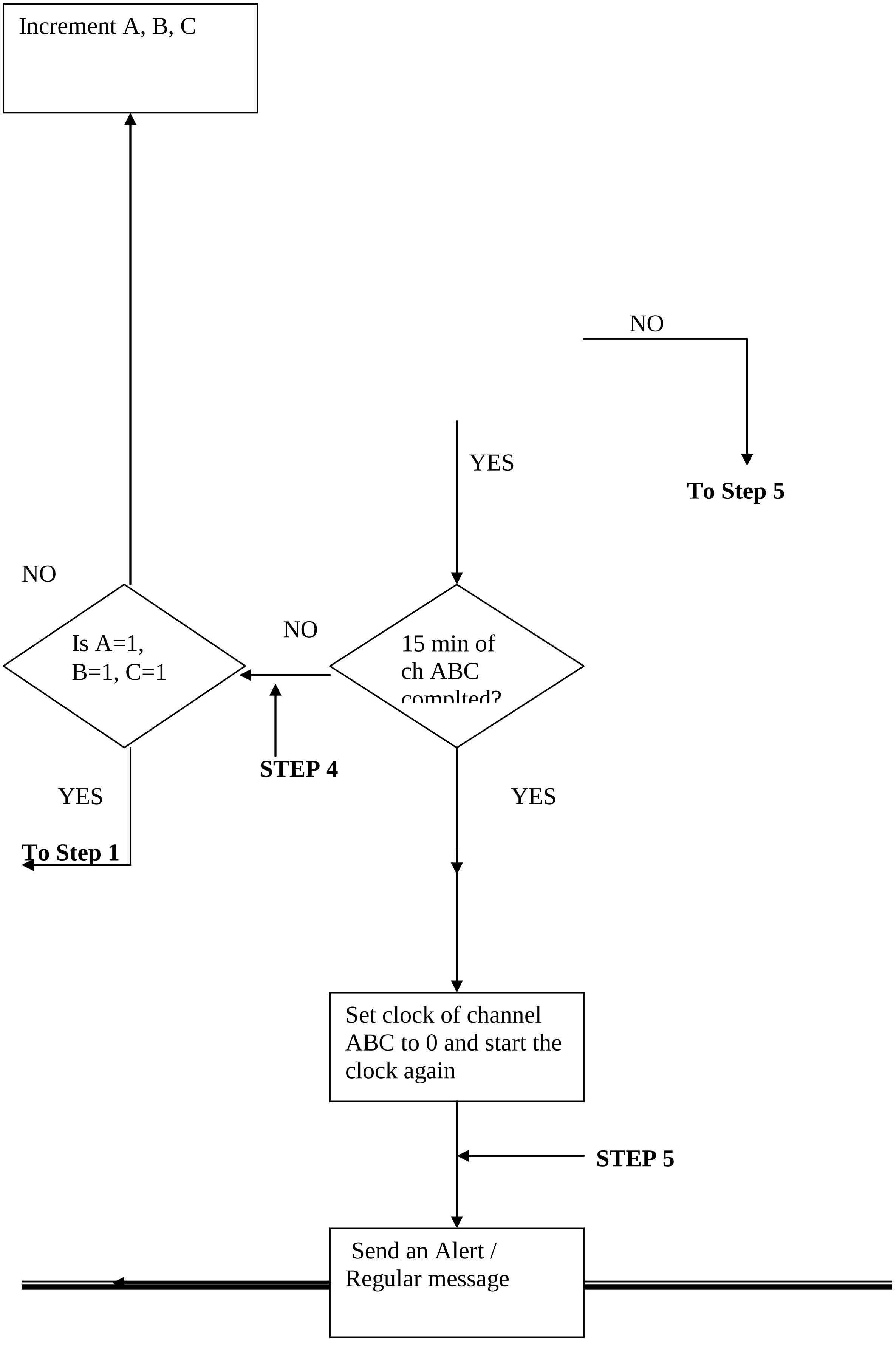

Increment A, B, C
NO
To Step 5
YES
NO
Is A=1, B=1, C=1
NO
15 min of ch ABC complted?
STEP 4
YES
YES
To Step 1
Set clock of channel ABC to 0 and start the clock again
STEP 5
Send an Alert / Regular message

**To Step 4**

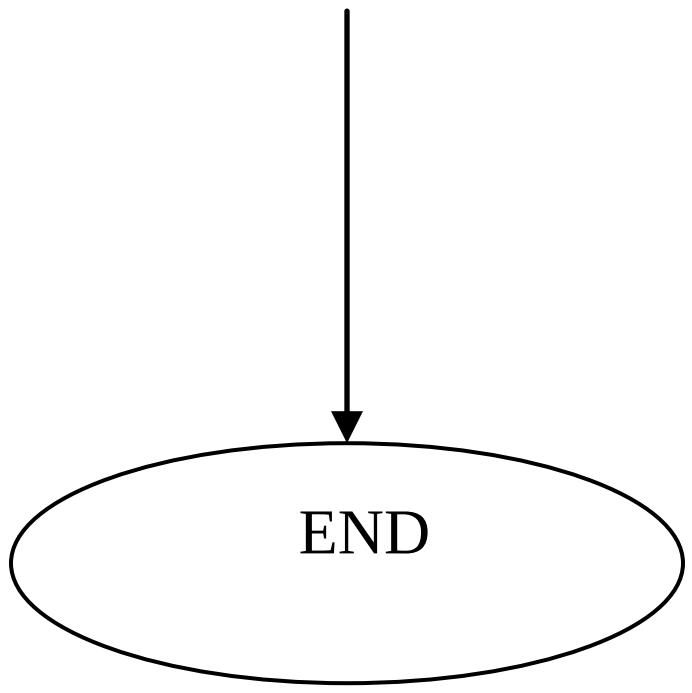

**RESULTS AND DISCUSSION**

## 7.1 RESULTS AND CONCLUSION:

This system could successfully send Short Messaging Services (SMS) to the concerned doctor's mobile phone as desired. The doctor could set the critical values for body temperature and Heart rate for individual patients in the monitoring computer terminal. The program was built to monitor two patients at a time and report the results to concerned doctors.

**Fig. 7.1 Front end of *Real time Patient's data monitoring system***

For the above entries if the temperature and heart rate of the two patients were within the critical limits, a regular SMS was sent to the concerned doctor after every 15 min.
The SMS sent to doctor1 (9844120647) who is concerned with patient of bed no. 1 reads as: "***The patient Ram Gopal Verma of bed no.: 1 has temperature 97.34 deg Fahrenheit & Heart rate 74***".
If there is any change in temperature or heart rate above or below the critical values, immediately an alert SMS is sent to the concerned doctor. An alert SMS repeats only after every minute if the condition persists. It would look like this:
**"*ALERT, the patient Ram Gopal Verma of bed no.: 1 has temperature 103.69 deg Fahrenheit & Heart rate 74*"**

This system is very helpful in large hospitals where the concerned doctor cannot be with the patient all the time. It saves the doctor's valuable time by constantly monitoring and sending real time parameter values to him at regular intervals of time.

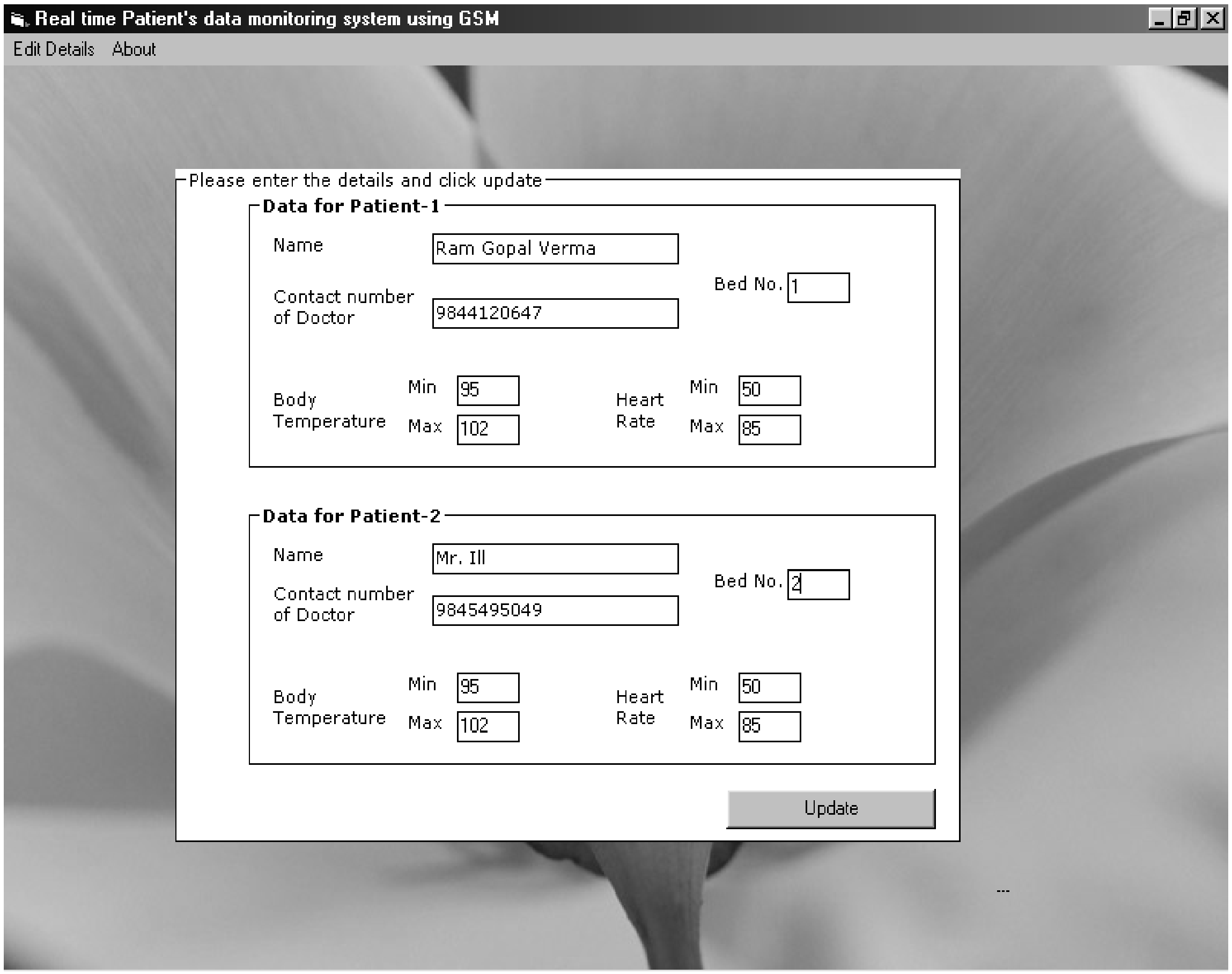

## 7.2 FUTURE SCOPE:

- Saved data can be represented in XML and could be indexed using SOLR API's
- Data can also be shared with other organizations using Open Archive Initiative.